# Research on the X-Ray Polarization Deconstruction Method Based on Hexagonal Convolutional Neural Network

Ya-Nan Li[1,2] · Jia-Huan Zhu[3] · Huai-Zhong Gao[1,2] · Hong Li[3] · Ji-Rong Cang[4] · Zhi Zeng[2] · Hua Feng[3] · Ming Zeng[1,2]

*1. Key Laboratory of Particle & Radiation Imaging (Tsinghua University), Ministry of Education, Beijing 100084, China*

*2. Department of Engineering Physics, Tsinghua University, Beijing 100084, China*

*3. Department of Astronomy, Tsinghua University, Beijing 100084, China*

*4. StarDetect Co., Ltd , Beijing 100084, China*

E-mail: zengming@tsinghua.edu.cn (Ming Zeng)

**Abstract:**

Track reconstruction algorithms are critical for polarization measurements. In addition to traditional moment-based track reconstruction approaches, convolutional neural networks (CNN) are a promising alternative. However, hexagonal grid track images in gas pixel detectors (GPD) for better anisotropy do not match the classical rectangle-based CNN, and converting the track images from hexagonal to square results in loss of information.
We developed a new hexagonal CNN algorithm for track reconstruction and polarization estimation in X-ray polarimeters, which was used to extract emission angles and absorption points from photoelectron track images and predict the uncertainty of the predicted emission angles. The simulated data of PolarLight test were used to train and test the hexagonal CNN models. For individual energies, the hexagonal CNN algorithm produced 15-30% improvements in modulation factor compared to moment analysis method for 100% polarized data, and its performance was comparable to rectangle-based CNN algorithm newly developed by IXPE team, but at a much less computational cost.

**Keywords**   X-ray polarization · Track reconstruction · Deep learning · Hexagonal conventional neural network

# 1  Introduction

Astronomical X-ray polarimetry is a powerful tool for probing the magnetic fields, geometries, and emission physics of high energy astrophysical sources [1-4]. Astronomical X-ray polarization measurements originated in the 1960s to detect the soft X-ray polarization of crab nebula, Scorpius X-1, and other objects using a Bragg diffraction polarimeter and Thomson scattering polarimeter [5-9]. However, limited by the sensitivity of the polarimeter, astronomical X-ray polarization measurements have been stalled for more than 40 years since the experiments on the OSO-8 satellite in 1968.

In the energy range of a few keV, the photoelectric effect dominates the light–matter interactions. The differential cross-section of photoelectrons is proportional to $\cos 2\varphi$; here, $\varphi = \varphi_e - \varphi_0$, where $\varphi_e$ is the azimuthal angle of the photoelectron and $\varphi_0$ is the X-Ray`s electric vector position angle (EVPA) [10]. Therefore, the polarization fraction and polarization angle EVPA of the X-ray source can be obtained by measuring the emission angles of numerous photoelectrons. With the development of micro-pattern gas detectors, polarimetry based on the photoelectric effect has become possible by measuring the emission angles of photoelectrons, greatly improving polarization sensitivity [11]. The PolarLight CubeSat test [12-13], IXPE mission [14], and scheduled eXTP mission [15-16] all use the gas pixel detectors (GPD) to detect polarization.

However, owing to Coulomb scattering, transversal diffusion during drift, and electronic noise, the reconstruction of emission angles from photoelectron tracks is complicated. The performance of the photoelectron track-reconstruction algorithm significantly affects the sensitivity of the polarimeter.

Recently, there have been two types of track reconstruction methods: traditional algorithms, which include the moment analysis method [17-18], adaptive cut method [19], and graph-based methods [20]; and reconstruction methods based on convolutional neural networks (CNN) [21-23], which demonstrate great advantages in track reconstruction owing to their powerful image processing capabilities. However, major space polarimetry missions use a hexagonal-pixel ASIC to read out track images for better isotropy [12-16], resulting in photoelectron track images with a hexagonal-pixel structure (Fig. 1). The existing CNN-based methods use a classical rectangle-based CNNs with an additional step to convert the hexagonal-pixel track images into approximate square-pixel track images, which results in loss of information in the photoelectron image.

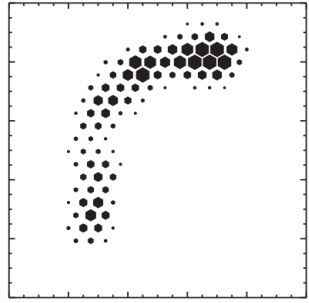

Fig. 1 Typical photoelectron track image in GPD

Therefore, developing CNN methods that match the hexagonal pixel track structure is a worthwhile research direction with good scientific significance and promising performance. Hexagonal CNNs are deep-learning models based on a hexagonal pixel structure. In hexagonal CNNs, hexagonal convolutional kernels are used instead of the rectangular convolutional kernels in classical CNNs to better capture the spatial context information in hexagonal pixel images. Using hexagonal CNNs to process hexagonal pixel photoelectron track images in the GPD is expected to achieve better polarization reconstruction.

In this study, we propose a new X-ray polarization reconstruction method based on hexagonal CNNs.

The remainder of the paper is structured as follows. Section 2 briefly introduces the hexagonal CNNs and uncertainty quantification in deep learning. Section 3 describes the training procedure for the hexagonal CNNs in photoelectron track reconstruction. Section 4 presents the prediction and reconstruction results of the hexagonal CNN method. Finally, we conclude the paper and present future development prospects in Section 5.

## 2  Hexagonal CNNs and the uncertainty quantification in deep learning

### 2.1  Hexagonal CNNs

CNNs have received considerable attention in recent years because of their excellent performance in computer vision and big data [24-26]. With the increase in application fields, classical CNN based on the Cartesian architecture can no longer meet the demands of more complex problems. Many studies have made significant advances in the design of network architectures and convolutional operations, generalising CNN to multi-view [27], non-Euclidean spaces [28], and other domains.

Typically, images are acquired using square sensor arrays. However, square grids are not the best solutions for planar segmentation [29]. Compared with square grids, hexagonal grids have many advantages such as 6-fold rotational symmetry, smaller edge-to-area ratio, and equidistant neighbours. Hexagonal grids have been widely used in cosmological, astrophysical, and visual systems, such as ground-based ray observations with Imaging Atmospheric Cherenkov Telescopes (IACTs) [30], the Cherenkov Telescope Array (CTA) [31], IceCube [32], and PolarLight test.

A hexagonal CNN is a class of deep learning based on hexagonal grids, in which the hexagonal convolution kernel is used to replace the rectangular convolution kernel in the classical CNN. The differences between the two types of convolution kernels are shown in Fig. 2. Compared with classical CNN, hexagonal CNN exhibit unique advantages in the fields of aerial scenes and geospatial information [33-34].

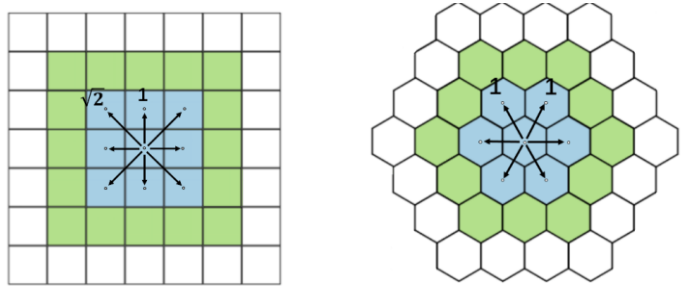

Fig. 2 Rectangular convolution kernel (left) and hexagonal convolution (right)

Despite the abovementioned advantages over classical CNNs, hexagonal CNNs have higher computational complexity and are generally more difficult to train. Existing research on hexagonal CNNs involves two main approaches. One is to implement hexagonal convolution by reusing existing highly optimised rectangle-shaped convolution routines, such as HexagDLy [35] and HexagonNet [36], whereas other studies focus on native hexagonal CNN architectures which can implement hexagonal convolutional operations directly, such as HexCNN [37]. While native hexagonal CNNs offer tremendous advantages in terms of training time and memory space cost, they cannot yet be implemented on GPUs and take advantage of the efficient parallel computation of GPUs to accelerate model training and inference [37]. HexagDLy is a Python library that performs convolution and pooling operations for hexagonal pixel data. Fig. 3 demonstrates the convolutional implementation with a hexagonal kernel of size one in HexagDLy as an example [35]. We constructed a hexagonal CNN architecture for track reconstruction using HexagDLy, considering its flexibility and user friendliness. The detailed hexagonal CNN architecture for photoelectron track reconstruction is discussed in Section 3.

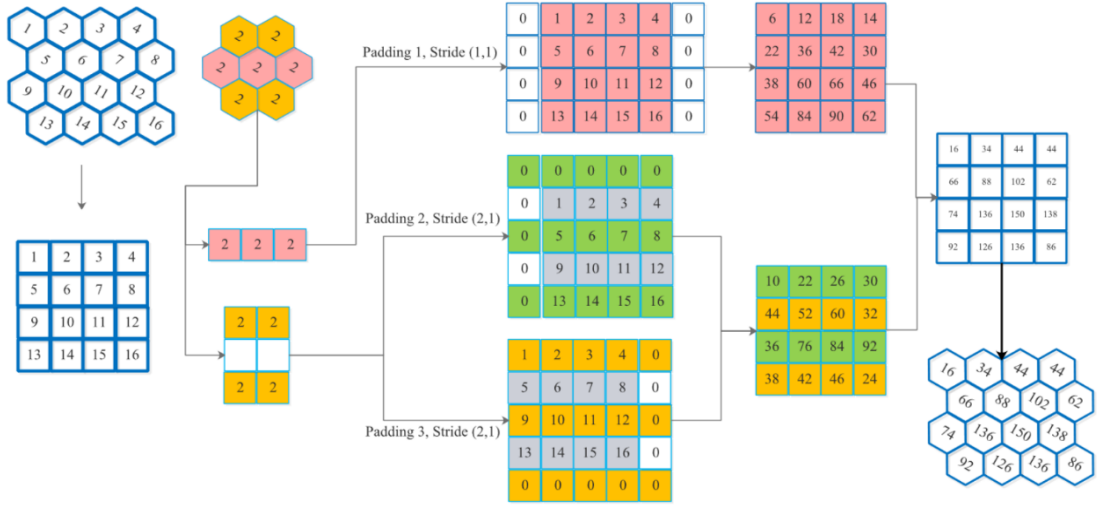

Fig. 3 An example of hexagonal convolution with a kernel of size 1 in HexagDLy

## 2.2 Uncertainty quantification

Uncertainty estimation of deep learning is also a popular research direction, which allows a neural network to output not only a prediction $\hat{y}$ for an input x, but also the predictive uncertainty $\sigma_{\hat{y}}$, greatly expanding the application fields of deep learning. Uncertainty estimation is also helpful for X-ray polarization reconstruction, which was confirmed in [22-23].

There are two main types of uncertainties in deep learning: aleatoric uncertainty (also known as data uncertainty) and epistemic uncertainty (also known as model uncertainty) [38]. Aleatoric uncertainty is used to assess the uncertainty of data that arises because of class overlap or inherent noise in the data and cannot be reduced by collecting more data. Epistemic uncertainty is used to assess the model uncertainty caused by a lack of cognition regarding the distribution of the data or inadequate model structure. Theoretically, epistemic uncertainty can be reduced using more complex models, expanding data, or regularisation techniques [39].

The aleatoric uncertainty can be modelled by augmenting the loss function. For example, assuming that the noise of data obeys Gaussian distribution (i.e., $\varepsilon \sim N(0, \sigma^2)$), the predicted output of the model for a given input x is $y \sim N(\hat{y}, \sigma^2)$. The loss function that predicts aleatoric uncertainty $\sigma_a$ can be obtained by minimising the negative log-likelihood (NLL) loss function on all training samples (seen in Eq. 1).

$$L(y_i|x_i) = \frac{\log(\hat{\sigma}_a^2(x_i))}{2} + \frac{\|y_i - \hat{y}(x_i)\|_2^2}{2\hat{\sigma}_a^2(x_i)} \quad (1)$$

Epistemic uncertainty is significantly more difficult to quantify than aleatoric uncertainty, and many methods have been proposed to quantify epistemic uncertainty, including Bayesian Neural Networks (BNNs) [40], deep ensemble [41], and Evidential Deep Regression (EDR) [42].

BNNs introduce priori assumptions to model epistemic uncertainty by setting a prior distribution $\omega$ upon the weight parameters of neural network and using the dataset D to derive the posterior distribution $P(\omega|D)$ of $\omega$. However, BNNs are difficult to apply in practice because the posterior distribution $P(\omega|D)$ is usually intractable. Some BNNs' approximation algorithms have been proposed to estimate the epistemic uncertainty, such as Variational Inference [43] and Monte Carlo dropouts [44].

Deep ensembles are another powerful approach for modelling epistemic uncertainty and have been widely used in many applications. A deep ensemble uses an ensemble with several base models and can generate multiple predictions $\{\hat{y}_j\}_{j=1}^{M}$ for the same input x, where M is the number of models in the deep ensemble. The variance in the predictions can be used as an epistemic uncertainty. Deep ensembles are easy to implement and can achieve as good or better uncertainty estimation than approximate BNNs algorithms.

Furthermore, EDR directly learns the higher-order distribution of the neural network output. It uses a deterministic network to learn both aleatoric and epistemic uncertainties by placing evidential priors over the original loss function that predicts the aleatoric uncertainty. The EDR has achieved good results in many applications. However, a recent study found that EDR has theoretical shortcomings in terms of mathematical foundations [45].

Taking all these considerations into account, we chose to estimate the aleatoric uncertainty by augmenting the loss function and epistemic uncertainty using a deep ensemble which is more stable and easier to implement.

## 3 Hexagonal CNN model training for photoelectron track reconstruction

X-ray polarization reconstruction algorithms are typically divided into two steps. Firstly, extracting track features from individual photoelectron track images (also called track reconstruction), which typically include the photoelectron emission angles $(\varphi)$, absorption points $(x, y)$, and photoelectron energies $(E)$. Subsequently, the polarization parameters (polarization fraction and EVPA) of the X-ray source are estimated based on the emission angles extracted from a large number of photoelectrons. Among them, the extraction of track features from blurred photoelectron track images is the key to polarization reconstruction. The following section describes the implementation of the hexagonal CNN–based photoelectron track feature reconstruction algorithm in detail.

### 3.1 Dataset

Supervised learning was used for photoelectron track reconstruction. Considering that the true emission angle $\varphi$ in the experimental data is unknown following a distribution of $\cos 2\varphi$, the dataset used for track reconstruction needs to be generated by simulation.

PolarLight is a small X-ray GPD polarimeter onboard a CubeSat that has conducted on-orbit scientific observations. An ASIC designed by the INFN-Pisa group with a hexagonal pixel size of $50~\mu m$ and a pixel matrix of $352 \times 300$ (105k pixels) is used for track readout in PolarLight [46]. The PolarLight test has a complete simulation algorithm using the Monte Carlo Geant4/Garfield simulation and has passed consistency validation with the experimental data. The simulation algorithm was used to generate a photoelectron track dataset.

The photoelectron track features include the emission angles, absorption points, and photoelectron energies. Because the reconstruction of photoelectron energies is relatively simple and can be done well by non-CNN algorithms, we only reconstruct $(\varphi, x, y)$ in this paper.

The dataset used for CNN model training should be uniformly distributed; otherwise, the model may suffer from overfitting, low prediction accuracy, or biased prediction results. Hence, the parameters of the incident X-rays in the simulation algorithms were set considering the following:

1) To ensure a uniform distribution of emission angles, the polarization of the incident X-rays was set to 0. In other words, the incident X-rays are unpolarized.

2) To ensure that the hexagonal CNN model performed well in tracking feature extraction for the entire detector plane, the coordinates $(x, y)$ of the incident X-rays were uniformly distributed.

3) Considering that the effective energy range of PolarLight is 2-8 keV, the incident X-rays were uniformly distributed in the range of 2–9 keV, to ensure that the hexagonal CNN model was adequately trained for data at the edge of the energy interval. Because low-energy photoelectron track images are noisy and it is difficult to extract track features from them, the dataset is not expanded here for lower energies.

We simulated 870,050 photoelectron tracks with a uniform distribution of emission angles, absorption points, and energies and then split them into a training set (90%), a validation set (5%), and a test set (5%).

It is important to note that the photoelectron tracks are generated not only by photons interacting with the gas in the GPD but also by photons interacting with the detector components outside the gas volume (e.g. the beryllium window and the gas electron multiplier (GEM)), in which case the photoelectrons lose some of their energy and result in a low-energy tail in the energy histogram [22-23]. It is often difficult to recover the emission angles of the tail tracks from these tracks. Our study focused on reconstructing the photoelectron track generated within the GPD gas volume; therefore, these tail tracks were removed from the dataset.

In addition, because the HexagDLy used in this study is based on rectangle-shaped convolution, an image pre-processing is required, in which the hexagonal grid image is converted to a square grid image. Fig. 4 shows an example of the image pre-processing. In order to take full advantage of the 6-fold rotational symmetry of the hexagonal grid image, each photoelectron track image is converted into three input images with the operations of unrotated, rotated 60°, and rotated -60°. Furthermore, considering the photoelectron track length and the pixel size of the readout ASIC, we set the photoelectron track image size to $64 \times 64$.

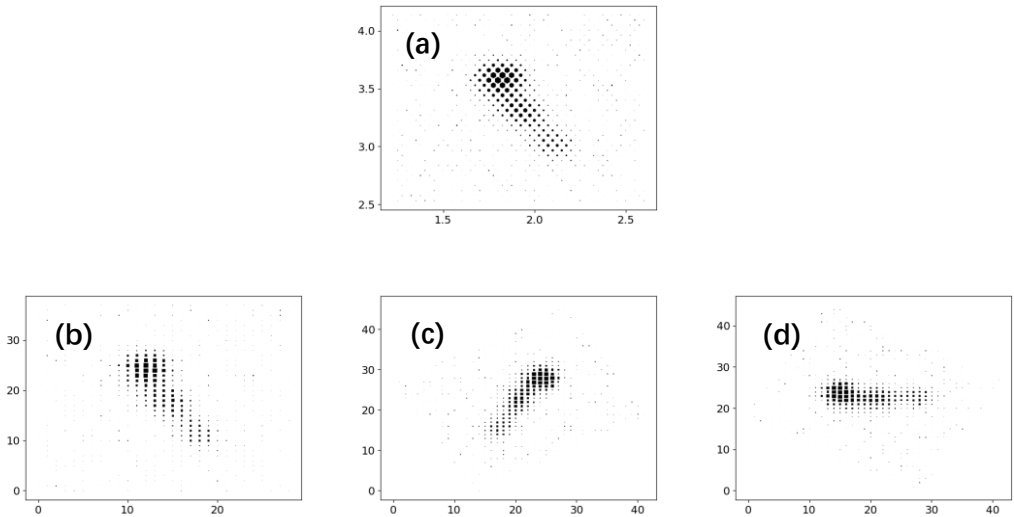

Fig. 4 Track image pre-processing. (a) original track image generated by the simulation algorithms; (b)–(d) input images of hexagonal CNN after pre-processing, unrotated (left), rotated 60° (middle), rotated -60° (right)

## 3.2 Loss function

Loss functions were used to evaluate how well the machine learning modelled the datasets. Considering that track reconstruction is a multitasking problem, it is necessary to establish loss functions for emission angles and absorption points separately.

The emission angles of photoelectrons are periodic and its distribution is more consistent with the Von Mises distribution, which is a continuous probability distribution with a range of 0 to $2\pi$ and is the circular analogue of the normal distribution on a line. To predict the epistemic uncertainty, the NLL of the Von Mises distribution is a better choice than the Gaussian NLL described in Section 2.2. The loss function of the emission angles based on the Von Mises distribution of a single hexagonal CNN model is shown in Eq. 2, with a detailed description in [23].

$$L_\varphi(v|x) = -\hat{\kappa}^a(\hat{v}_2 \cdot v_2) + \log I_0(\hat{\kappa}^a) \quad (2)$$

where $I_0$ is a modified Bessel function of the first kind with order 0, $v_2 = (\cos 2\varphi, \sin 2\varphi)$ considering that the polarization of X-ray is only associated with $2\varphi$, and non-negative $\hat{\kappa}^a$ is the predicted aleatoric VM uncertainty parameter of the emission angle. The circular variance of VM distribution can be derived from $\hat{\kappa}^a$ by Eq. 3 where $I_1$ is a modified Bessel function of the first kind with an order of 1.

$$\sigma_a^2 = 1 - \frac{I_1(\hat{\kappa}^a)}{I_0(\hat{\kappa}^a)} \quad (3)$$

Assuming that the epistemic VM uncertainty $\kappa^e$ also follows a Von Mises distribution VM(0, $\kappa^e$), $\hat{\kappa}^e$ can be estimated from the emission angles of a deep ensemble of M hexagonal CNN models [23].

$$\bar{R}^2 = (\frac{1}{M}\sum_{j=1}^{M} \cos 2\hat{\varphi}_j)^2 + (\frac{1}{M}\sum_{j=1}^{M} \sin 2\hat{\varphi}_j)^2 \quad (4)$$

$$\frac{I_1(\hat{\kappa}^e)}{I_0(\hat{\kappa}^e)} = \bar{R} \quad (5)$$

The total uncertainty variance $\sigma^2$ of the emission angle $\varphi$ can be obtained by summing the aleatoric uncertainty variance $\sigma_a^2$ and the epistemic uncertainty variance $\sigma_e^2$, i.e., $\sigma^2 = \sigma_a^2 + \sigma_e^2$. The total error $\sigma_i$ of emission angle prediction $\varphi_i$ for track $x_i$ is then given by Eq. 6.

$$\sigma_i = \frac{1}{2}\sqrt{\frac{1}{M}\sum_{j=1}^{M}(1 - \frac{I_1(\hat{\kappa}_{ij}^a)}{I_0(\hat{\kappa}_{ij}^a)}) + (1 - \frac{I_1(\hat{\kappa}_i^e)}{I_0(\hat{\kappa}_i^e)})} \quad (6)$$

where a factor of 1/2 is used to transform the errors from $2\varphi_i$ to $\varphi_i$.

The loss function of the absorption points was the L2 loss function (Eq. 7) which is commonly used in CNNs. The uncertainty in the absorption points is not the focus of this study and can be obtained using Eq. 1 combined with a deep ensemble, if needed.

$$L_{xy}(x_0, y_0|\mathrm{x}) = \frac{1}{2}\|(x_0, y_0) - (\hat{x}(\mathrm{x}), \hat{y}(\mathrm{x}))\|_2^2 \quad (7)$$

The total loss function of a single hexagonal CNN model for track reconstruction is:

$$L = L_\varphi + \alpha\|\mathrm{v}_1 - \hat{\mathrm{v}}_1\|_2^2 + \beta L_{xy} + \sigma\|\omega\|_2 \quad (8)$$

where $\|\mathrm{v}_1 - \hat{\mathrm{v}}_1\|_2^2$ is added to make the hexagonal CNN model can predict emission angles in a range of $2\pi$; the last term $\sigma\|\omega\|_2$ is used to prevent training overfitting.

These individual loss functions are connected by three hyperparameter weights $\alpha, \beta, \sigma$. An individual hexagonal CNN model will predict a 5-dimensional vector $(\cos\varphi, \sin\varphi, \kappa, x, y)$ for a given track image input x.

## 3.3 Hexagonal CNN Architecture

The reconstruction of emission angles and absorption points is complex, and a simple architecture with 3 or 4 convolution layers cannot satisfy the demand for track reconstruction. Considering the photoelectron track-image features, we built hexagonal CNNs for track reconstruction based on the ResNet-18 architecture [47].

Residual block is the basic unit of a residual network. A hexagonal residual block (Fig. 5) was built using hexagonal convolution operation provided by HexagDLy. It has a hexagonal convolution layer with a kernel of size 1 defined by HexagDLy, a batch normalisation layer, and a ReLU activation function, followed by another hexagonal convolution layer with a kernel of size 1 and a batch normalisation layer. Subsequently, a skip connection skips these layers and directly adds a ReLU activation function at the end. These hexagonal residual blocks are repeated to form a complete hexagonal CNN architecture for track reconstruction.

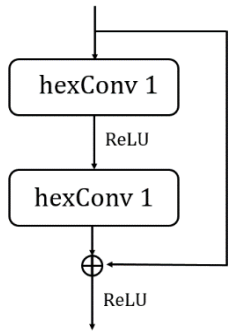

Fig. 5 Residual block structure based on hexagonal convolutional layers

In this hexagonal CNN architecture (Table 1), the conv1 layer uses a hexagonal convolution layer (kernel size = 1) and a hexagonal maximum pooling layer (kernel size = 1, stride = 2) to extract track features. Then conv2-conv5 layers formed by hexagonal residual blocks were used to extract deeper track features. Finally, feature maps generated by conv5 were converted into a 5-dimensional vector $(\cos\varphi, \sin\varphi, \kappa, x, y)$ using an average pooling layer and a fully connected layer.

Table 1: Hexagonal CNN for photoelectron track reconstruction

| layer name | layer | output size | feature map |
|---|---|---|---|
| conv1 | hexconv 1<br>hexMaxPool2d | $32 \times 32$ | 64 |
| conv2 | Residual block×2 | $16 \times 16$ | 64 |
| conv3 | Residual block×2 | $8 \times 8$ | 128 |
| conv4 | Residual block×2 | $4 \times 4$ | 256 |
| conv5 | Residual block×2 | $2 \times 2$ | 512 |
| avgpool | Average pooling | $1 \times 1$ | 512 |
| fc | Fully connected | 5 | |

### 3.4 Training

A standardisation operation was applied to the training data before training to prevent vanishing and exploding gradients and to speed up convergence.

$$x_{norm} = \frac{x - \mu}{\sigma} \quad (9)$$

where $\mu$ is the pixel mean and $\sigma$ is the pixel standard deviation, calculated over the entire training set of track images.

The hexagonal CNN model was optimised using stochastic gradient descent with momentum (SGD), which is a typical optimisation algorithm used in deep learning [48]. The learning rate decayed in steps starting at 0.005. The model parameters were randomly initialised before training to provide a different start for training and generate 5 different initialised hexagonal CNN models for deep ensembles. Considering the memory consumption of the hexagonal CNN model, batch sizes of 512 and 1024 were selected. The hexagonal CNN model training lasts for 150 epochs and the hyperparameters in the loss function were chosen for $\alpha = 0.3, \beta = 0.2, \sigma = 5 \times 10^{-5}$.

## 4 Results

This section demonstrates the performance of the hexagonal CNN method for track and polarization reconstructions on simulated PolarLight track images.

### 4.1 Emission angles reconstruction and uncertainty estimation

The reconstruction of photoelectron emission angles is the basis of X-ray polarization reconstruction, and a more accurate emission angle reconstruction can greatly improve the performance of the polarization reconstruction algorithm.

The hexagonal CNN method of the paper uses $(\cos\varphi, \sin\varphi)$ to predict the emission angles $\varphi$. As described in Section 3, image rotation augmentation was used in this study to convert a tracked image into three input images. Therefore, each hexagonal CNN model outputs three sets of predicted vectors for each track image. Considering that a deep ensemble method is used to obtain the epistemic uncertainty by estimating the predictions of M (M=5 in this study) hexagonal CNN models, there are 3M predictions for each track image. The emission angle for a single track predicted by the hexagonal CNN method was calculated using Eq.10:

$$\bar{\varphi}_i = \arctan2(\frac{1}{3M}\sum_j^{3M}\sin\varphi_{ij}, \frac{1}{3M}\sum_j^{3M}\cos\varphi_{ij}) \quad (10)$$

Fig. 6 shows the performance of the emission angle reconstruction for the moment analysis and the hexagonal CNN method with photoelectrons energies of 3 keV and 9 keV. Because of the inability to accurately distinguish between the beginning and end of a photoelectron track sometimes, especially at lower energies, there is a 180° confusion in the emission angle reconstruction, which is shown as two sub-bright lines parallel to the central bright line in Fig. 6. Notably, the 180° confusion does not affect polarization reconstruction, where the EVPA ranges from $-\pi/2$ to $\pi/2$. It can be seen that compared to the moment analysis, hexagonal CNN method distinguishes the beginning and the end of tracks with higher accuracy and have a higher emission angle reconstruction accuracy.

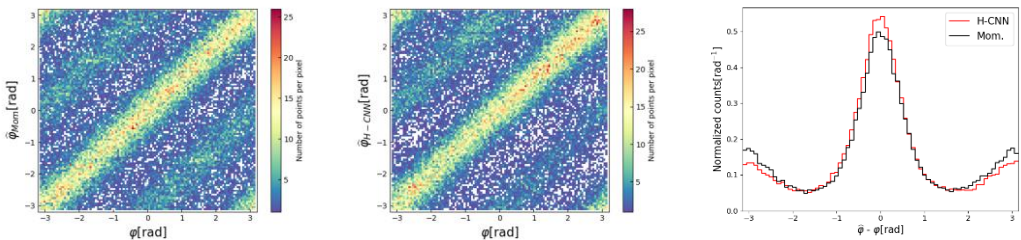

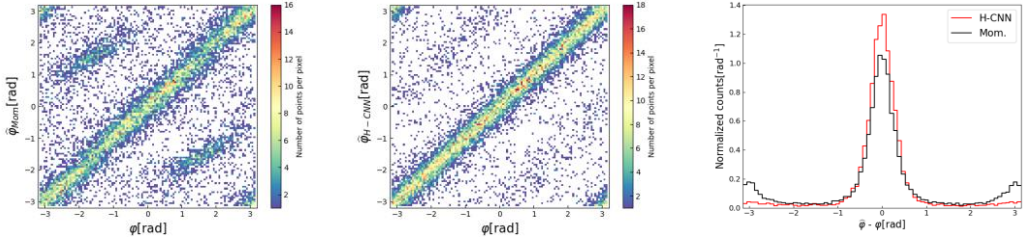

Fig. 6 Emission angle reconstruction for the moment analysis (left) and the hexagonal CNN method (middle) and histograms (right) of the differences between the predicted emission angle $\hat{\varphi}$ and the true emission angle $\varphi$ for moment analysis (Mom., black) and hexagonal CNN method (H-CNN, red) with photoelectrons energies of 3 keV. (top) and 9 keV (bottom)

The root mean square error (RMSE) measured by Eq. 11 was also used to evaluate the accuracy of the emission angle reconstruction. Fig. 7 demonstrates the RMSE of the emission angles on an unpolarized PolarLight dataset as a function of true energy for both the moment analysis and hexagonal CNN method.

$$\text{RMSE} = \sqrt{\frac{1}{N}\sum_{i}^{N}(\hat{\varphi}_i - \varphi_i)^2} \quad (11)$$

where $\hat{\varphi} - \varphi$ is collapsed into the range $-\frac{1}{2\pi} \sim \frac{1}{2\pi}$ because the polarization reconstruction only depends on $2\varphi$.

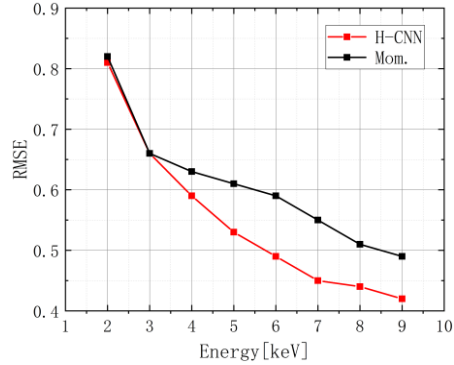

Fig. 7 RMSE of emission angle reconstruction for both the moment analysis (black) and hexagonal CNN method (red)

The hexagonal CNN method did not significantly improve the reconstruction accuracy of the emission angles compared to the moment analysis for low-energy photoelectrons with short and noisy tracks. As the X-ray energy increases, the photoelectron tracks become longer with clearer beginning segments. Both reconstruction algorithms provide high accuracy for emission angle reconstruction, and the hexagonal CNN method is significantly better than the moment analysis for these complex tracks.

Another important prediction of emission angle is the predicted error $\sigma_i$ which can be calculated by the output κ of hexagonal CNN method (Eq.6). The predicted uncertainty in the emission angle can help screen out events with poor emission angle reconstruction, thus improving the effectiveness of the X-ray polarization estimation. Since it is difficult to determine quantitatively whether the predicted error of the hexagonal CNN method is correct, we perform a basic calibration of the reasonableness by comparing predicted error with the factors affecting of the effectiveness of the emission angle reconstruction.

The difficulty in reconstructing the emission angle of the photoelectron track is related to the degree of transversal electron diffusion during drift, degree of track shortening owing to the projection of the 3D photoelectron track onto a 2D readout plane, and photoelectron energy. To facilitate this discussion, a coordinate system was established for the effective sensitive volume of the PolarLight GPD. The effective sensitive volume of PolarLight is $15 \text{ mm} \times 15 \text{ mm} \times 10 \text{ mm}$. We define the x-y plane as the plane of the readout, the z-axis direction as the direction of the electric field, and the coordinates of the centre point of the effective sensitive volume as (0,0,0).

It is visualised of the distributions of predicted error $\sigma_i$ given by hexagonal CNN as functions of these factors in Fig. 8. Fig. 8 (a) shows the distribution of the predicted error as a function of the drift distance, d, which is defined as the distance between the absorption point of the photon and the GEM in the GPD. The degree of the transversal electron diffusion during drift is related to d, with a standard deviation $\sigma_{\text{drift}} = \sigma_{\text{f}}\sqrt{d}$, where $\sigma_{\text{f}}$ is the diffusion coefficient of GPD gas. As the drift distance increased, the transversal diffusion became more severe, and the predicted error of the emission angles was higher. Fig. 8 (b) shows the distribution of the predicted error as a function of the photoelectron scattering angle, which is defined as the angle between the directions of the incident X-rays and photoelectrons. The more parallel the photoelectron direction is to the readout plane, the longer is the track projected onto the readout plane, resulting in a smaller predicted error in the emission angles. Fig. 8 (c) shows the distribution of the predicted error as a function of photoelectron energy. As the photoelectron energy increased, the photoelectron track became longer with clearer beginning segments, and the predicted error of the emission angle decreased.

Taking the above discussion together, the predicted error of the emission angle obtained by the hexagonal CNN method is reasonable, reflecting the degree of blurring of the photoelectron tracks and the difficulty of reconstructing the emission angle.

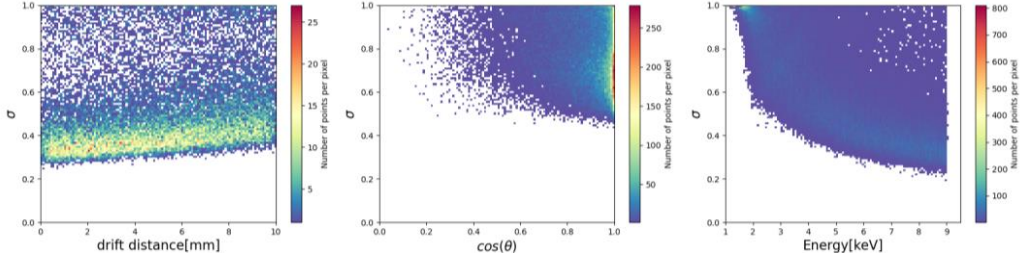

Fig. 8 Distribution of predicted error as function of the drift distance (left), photoelectron scattering angle (middle), and photoelectron energy (right)

Furthermore, the relationship between the real emission angle reconstruction error $\hat{\varphi} - \varphi$ and the predicted error is shown in Fig. 9. It can be seen that, as expected, the larger the real emission angle reconstruction error, the higher the uncertainty in its predicted emission angle.

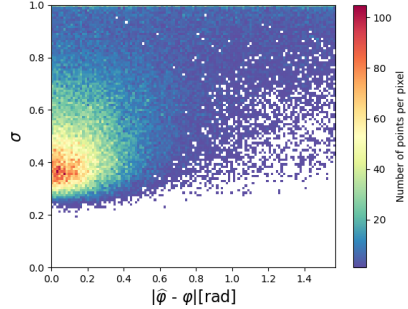

Fig. 9 Relationship between the real emission angle reconstruction error $\hat{\varphi} - \varphi$ and predicted error

### 4.2 Absorption points reconstruction

Reconstruction of the absorption points is important for improving the spatial resolution of the polarimeter. The absorption point accuracy can be evaluated using the half-power diameter (HPD), a commonly used parameter in X-ray imaging that is defined as the diameter of a circle that can cover exactly 50% of the reconstructed absorption point, taking the true absorption point as the centre of the circle. Therefore, the larger the HPD, the worse the reconstruction of the photoelectric absorption point. Fig. 10 compares the absorption point accuracy of the moment analysis and hexagonal CNN method for photoelectron tracks at different energies. It can be seen that the hexagonal CNN method is better than moment analysis method, especially for highly complex photoelectron tracks at high energies.

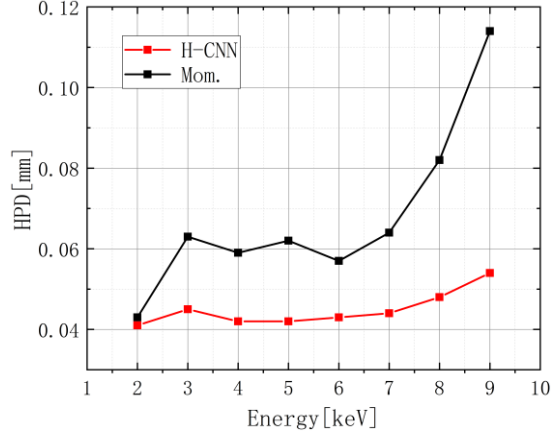

Fig. 10 Absorption points reconstruction with the moment analysis (black) and hexagonal CNN method (red)

## 4.3 Polarization estimation

The polarization reconstruction performance of the polarization estimation algorithm directly affects the polarimeter sensitivity. We analysed the polarization reconstruction performance of the hexagonal CNN method and compared it with the moment analysis and the rectangle-based CNN method developed by IXPE team.

We generated polarized and unpolarized simulation tracks using PolarLight simulation algorithms for polarization reconstruction analysis.

The binned modulation curves created using the predicted emission angles for the unpolarized and 100% polarized simulated data are shown in Fig. 11. The residual systematic modulation curve of the hexagonal CNN method is as flat as that of the moment analysis, indicating that hexagonal CNN method does not introduce redundant systematic errors. In addition, hexagonal CNN method recovers significantly more modulation in the polarized data compared to the moment analysis. An unbinned polarization estimation algorithm based on the Stokes parameters was used to estimate the polarization fraction and EVPA from a set of predicted track angles. Fig. 12 shows the recovered modulation response on the simulated PolarLight data for moment analysis, rectangle-based CNN method developed by IXPE team, and our hexagonal CNN method. It can be seen that our hexagonal CNN method performs better than the moment analysis, with 15–30% improvements in the modulation factor for individual energies.

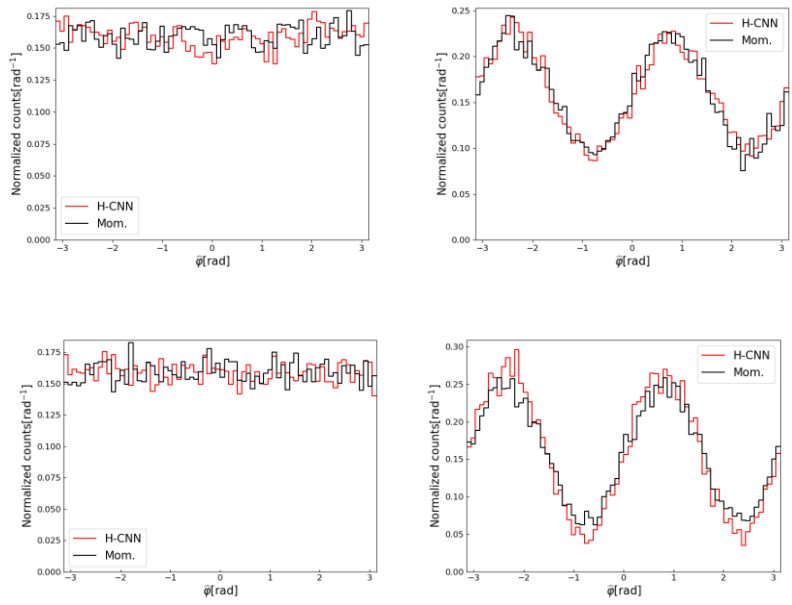

Fig. 11 Track angle reconstruction for unpolarized (left column) and polarized (right column) simulated data for 3 and 9 keV with moment analysis method (black) and hexagonal CNN method (red)

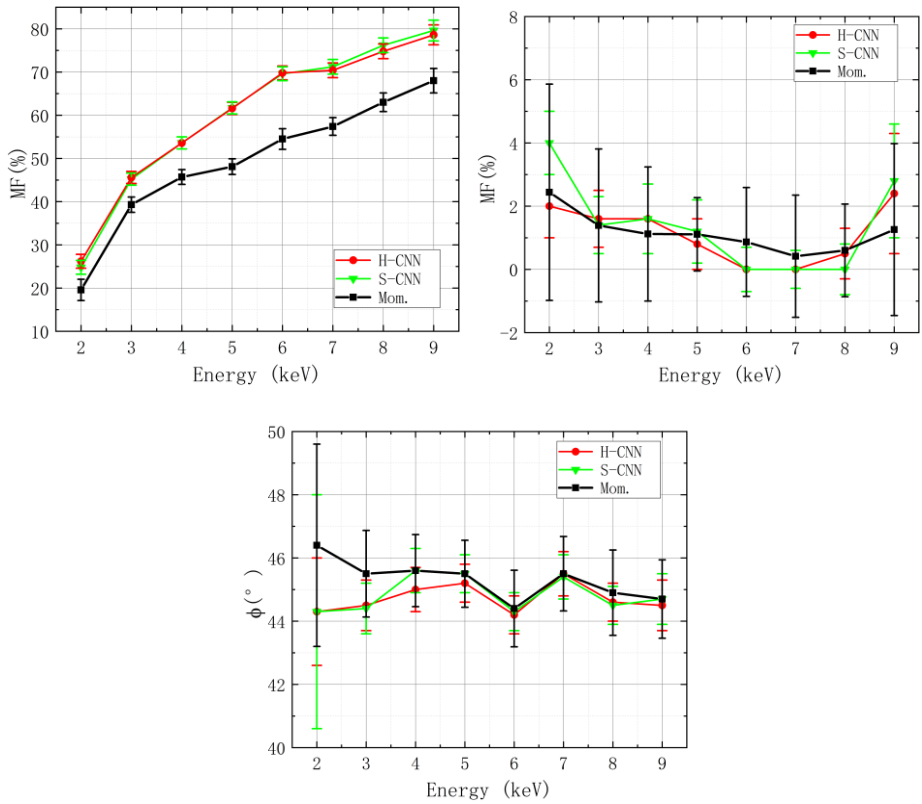

Fig. 12 Modulation response of for moment analysis (black), rectangle-based CNN method developed by IXPE team (S-CNN, green), and our hexagonal CNN method (red). (a) Response on a 100% polarized data. (b) Response on the unpolarized data. (c) Recovered

EVPA for a 100% polarized data

Compared to the CNN method developed by IXPE team based on classical rectangular convolution, our hexagonal CNN method has a similar performance in polarization reconstruction, although the hexagonal convolutional structure of hexagonal CNN is better matched with hexagonal sampling tracks. This may be because the double-channel input track images of the CNNs compensate for the loss during the conversion from hexagonal images to square images or because the existing neural network method is already close to the upper limit of polarization reconstruction owing to the blurring of the photoelectron tracks, which is difficult to improve with a better CNN architecture.

The advantage of the hexagonal CNN method is that each track image is converted into three single-channel input images for prediction, whereas the CNN method developed by IXPE team is converted into three double-channel input images, thereby halving the amount of input data for the hexagonal CNN. However, the existing hexagonal convolution is mainly implemented based on rectangular convolution; therefore, the memory of the hexagonal CNN is higher, which can be improved using native hexagonal CNN architectures.

## 5 Conclusion

We developed a track reconstruction and polarization estimation algorithm based on hexagonal CNNs to match the hexagonal grid tracks in the GPD for X-ray polarization measurement. The emission angles, absorption points, and uncertainties of the emission angles of X-ray photoelectron tracks were predicted using the hexagonal CNN method developed in this study. The predicted absorption points were used for image reconstruction, while the predicted emission angles and prediction uncertainties were used to estimate the polarization of the X-ray source. We tested the proposed hexagonal CNN method using simulated PolarLight data. The result shows that the performance of absorption points reconstruction in HPD by the hexagonal CNN method is better than moment analysis method, and the modulation factor of the hexagonal CNN method produces 15-30% improvements compared to the moment analysis method. Compared to the CNN reconstruction algorithm developed by IXPE team based on classical rectangular convolution, our hexagonal CNN method has comparable performance but with much less computational cost. Our hexagonal CNN method also provides a good research basis for the development of polarization reconstruction algorithms for the eXTP mission.